\documentclass[aps,prl,floats,twocolumn]{revtex4}
 \usepackage{graphicx}
\usepackage{amssymb}
\begin{document}
\title{Fluctuation-regularized Front Propagation Dynamics}

\author{Elisheva Cohen} 
\affiliation{Dept. of Physics, Bar-Ilan  University, Ramat-Gan, IL52900 Israel}
\author{David A. Kessler}
\affiliation{Dept. of Physics, Bar-Ilan University, Ramat-Gan, IL52900 Israel}
\author{Herbert Levine} 
\affiliation{Center for Theoretical Biological
  Physics, University of California, San Diego, 9500 Gilman Drive, La
  Jolla, CA 92093-0319} 
\date{\today}
\begin{abstract}
  We introduce and study a new class of fronts in finite particle
  number reaction-diffusion systems, corresponding to propagating up a
  reaction rate gradient. We show that these systems have no
  traditional mean-field limit, as the nature of the long-time front
  solution in the stochastic process differs essentially from that
  obtained by solving the mean-field deterministic reaction-diffusion
  equations. Instead, one can incorporate some aspects of the
  fluctuations via introducing a density cutoff.  Using this method,
  we derive analytic expressions for the front velocity dependence on
  bulk particle density and show self-consistently why this cutoff
  approach can get the correct leading-order physics.
\end{abstract}
\maketitle

Many physical, chemical, and biological systems exhibit fronts which
propagate through space. Familiar examples range from chemical
reaction dynamics such as flame fronts~\cite{kpp}, phase transitions
such as solidification~\cite{kkl}, the spatial spread of
infections~\cite{blumen}, and even the fixation of a beneficial allele
in a population~\cite{fisher}. In all these cases, the underlying
dynamics of individual constituents (from molecules to organisms)
gives rise multiple macroscopic states. It is thus
of great interest to understand the universality classes of fronts
which govern what will happen when systems such as these are prepared
in a spatially heterogeneous manner. These classes determine the
selection of propagation speed, the sensitivity to 
particle-number fluctuations, and the stability of the front with respect to
deviations from planarity.

To date, several different classes of fronts have been described.
Perhaps the simplest one, exemplified by the case of solidification,
is that wherein a thermodynamically stable phase replaces a metastable
one~\cite{kkl}. Here the mean-field front velocity is determined via
the requirement that there exists a heteroclinic trajectory of the
steady-state problem (obtained by assuming that all fields depend only
on the traveling coordinate $z \equiv x-vt$) connecting the metastable
phase at $+\infty$ with the stable one at $-\infty$.  This type of
front is robust with respect to fluctuations, with power-law
corrections in $1/N$ (where $N$ is the number of particles per site in
the final state) to the mean-field limit~\cite{kns}.  A second class
is exemplified by the simple infection model $A+B \rightarrow 2A$ on
1d lattice with equal $A$ and $B$ hopping rates~\cite{blumen};
 this process leads in
the mean-field limit to the well-known Fisher equation~\cite{fisher} $\phi
_t \ = \ r \phi (1 - \phi) +D \phi _{xx}$. Here propagation is into
the linearly unstable $\phi =0$ state and $\phi$ is just the number of
$A$ particles at a site, normalized by $N$. Recent
work~\cite{bd,kns,vansaarloos,pechenik} has shown that the front behavior in
the stochastic model does approach that of the Fisher equation, where
the velocity is selected by the (linear) marginal stability
criterion~\cite{ben-jacob} to be $2\sqrt{rD}$, albeit with an
anomalously long transient $O(1/t)$ and anomalously large fluctuation
corrections $O(1/\ln ^2 N)$. There are also some surprising findings in
regard to both front stability in the case of unequal $D$~\cite{nature}, 
and also the scaling properties of front
fluctuations~\cite{moro}. Finally, there are also fronts which are
determined by the nonlinear marginal stability criterion (for example,
propagation into a nonlinearly unstable but linearly marginal state)
which have properties intermediate to the previous two classes.

In this work, we introduce a new class of fronts corresponding to
propagation up a reaction-rate gradient.  We focus again on the $A+B
\rightarrow 2A$ reaction~\cite{blumen}, with no $A$ particles and an
initial mean number $N$ of B particles at all sites past some initial
$x_0$, but now assume that the reaction probability depends on spatial
position. The two situations we wish to consider are respectively the
absolute gradient and the quasistatic gradient
\begin{eqnarray}
r_a(x) & =& \text{max}(r_{\text min},r_0+ \alpha x) \ ,\nonumber \\ 
r_q(x) & = & \text{max}(r_{\text min},\tilde{r}_0 + \alpha (x-x_f)) 
\end{eqnarray}
where $x_f(t)$ is the instantaneous position of the front, which
we identify as $x_f =  \   \frac{1}{2}  \sum_{i=0} N_A (x_i) x_i / \sum_{i=0} N_A (x_i)$,
where on the lattice $x_i =ia$.
(The minimum value of $r$ is just there to ensure that the bulk
remains stable for all $x$ and plays no important role.) The
quasistatic problem should lead to a translation-invariant front with
fixed speed $v_q (\tilde{r}_0,\alpha )$, whereas the absolute
gradient will lead to an accelerating front. In the latter case, one
can imagine ignoring the acceleration, obtaining thereby an adiabatic
approximation to the velocity $v (t; r_0, \alpha ) \simeq v_q
(\tilde{r} _0 (t), \alpha )$ where $\tilde{r} _0 (t) = r_0 +\alpha
x_f (t)$. As we will see, fluctuation effects are absolutely
crucial as the naive mean-field theory gives rise to ``irregular"
fronts in a way which we will define shortly. It should be noted that
glimpses of this new class were obtained in studies of models of
Darwinian evolution~\cite{tsimring,evol1,rouzine}, but no general
understanding was attained.

In Fig. 1, we show numerical solutions of the mean field equation
(MFE, here just the Fisher equation with spatially varying $r$) versus
a stochastic stimulation~\cite{kns,stochastic} of the original continuous
time Markov-process, both for the absolute gradient case. The MFE
behaves {\em irregularly} in the sense that the front {\em never}
recovers from its initial conditions, i.e. the system never falls into
a dynamic attractor. Conversely, we define a {\em regular} front as
one for which changes in initial data (as long as $N_A$ remains zero
past some starting point) can only effect a time-translation of an
otherwise fixed front solution and hence would be invisible on a plot
of velocity versus position. Clearly, the stochastic process gives
rise to a regular front and thus cannot in any manner be approximated
by the MFE.

To get some insight into the notion of regularity, we employ a
heuristic approach in which we mimic the leading-order effect of
finite population number fluctuations by introducing a cutoff in the
MFE~\cite{mf-dla,kepler,tsimring}. 
This cutoff replaces $r(x)$ by zero if the
density $\phi$ falls below $k/N$ for some $O(1)$ constant $k$; this
change in the reaction term prevents the leading edge from spreading
too far, too fast. This idea has proven its reliability in the Fisher
system with {\em constant} reaction rate where it correctly predicts
the aforementioned anomalous effects~\cite{bd}.  A numerical
simulation of the cutoff equation reveals the recovery of regularity
(see Fig. 2a) in its long-time behavior; of course, the time it takes
to converge to the dynamic front attractor diverges as $N \rightarrow
\infty$ (data not shown). Returning to the simulation results in Fig
1b, we see that the cutoff MFE (using $k$ as a fitting parameter) does
a quantitatively accurate job of tracking the actual front dynamics.
Fig. 2b presents a comparison of the front profile from the cutoff
theory with that of the stochastic model; this was done for the
quasistatic model as the translation invariance facilitates the
necessary ensemble averaging.  Overall, it is quite remarkable how
well this simple approach works; later, we will use our analytic
approach to the cutoff MFE to give at least some new indications of
why this might be the
case.

\begin{figure}
  \includegraphics[width=.23\textwidth]{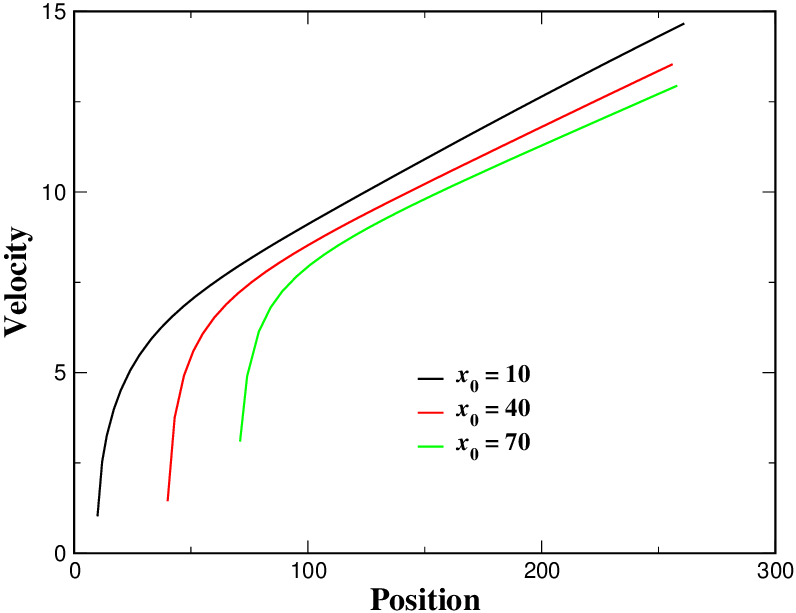}
  \includegraphics[width=.23\textwidth]{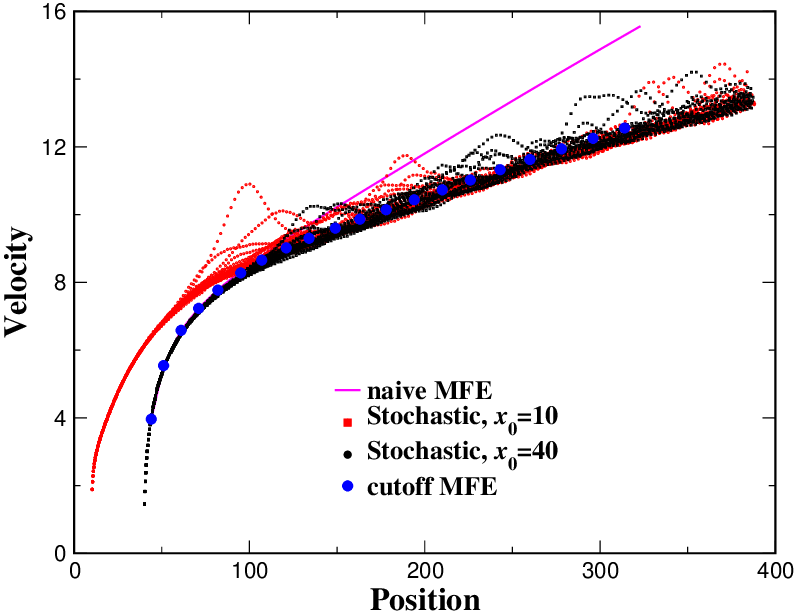}
\caption{Comparison of the MFE equation and the stochastic model, for $D=5$,
$r_0 =1$ and $\alpha =.02$. (a) Numerical integration of the MFE, with 
initial conditions $\phi =1$ for $x<x_0$, 0 otherwise. (b) Twenty 
simulations of the stochastic system with $N=10^{6}$, for each of 
two $x_0$'s, $x_0=10$, 40; also plotted is the corresponding cutoff MFE graph 
with $k=0.3$. We note in passing that the distribution of velocities seems 
highly skewed, with a large tail at velocities in excess of the mean. }
\label{fig1}
\end{figure}

We next turn to using the cutoff MFE to study the front velocity. In
Fig. 3, we present results for the front velocity at a fixed spatial
position for the stochastic model and the cutoff mean field theory as
a function of $N$. This is done both for the absolute gradient model
and for the corresponding quasistatic model. From the data, we can
draw the following conclusions. First, both models exhibit velocities
which increase without bound as a function of $N$; this is of course
radically different than what has been encountered in the previous
classes of propagating fronts. This behavior accounts for the fact
that the long-time dynamics never approaches that predicted by the
naive mean-field theory. Next, the two different models exhibit
similar velocities at small log $N$ but become increasingly different
as log $N$ goes to infinity. Finally, we note that at small enough log
$N$, the velocity can be approximated by just taking a cutoff version
of the usual Fisher equation result for a {\em fixed} reaction rate
$r_F = r_0 +\alpha \bar{x}$, i.e. completely neglecting the
reaction-rate gradient across the front. This can be explained by
noting that the effective interfacial width, the distance over which
the particle density drops from its bulk value $O(1)$ to its cutoff
value $O(1/N)$ scales as log $N$; hence one can neglect the gradient
if $\alpha \log{N}$ is small.
\begin{figure}
  \includegraphics[width=.23\textwidth]{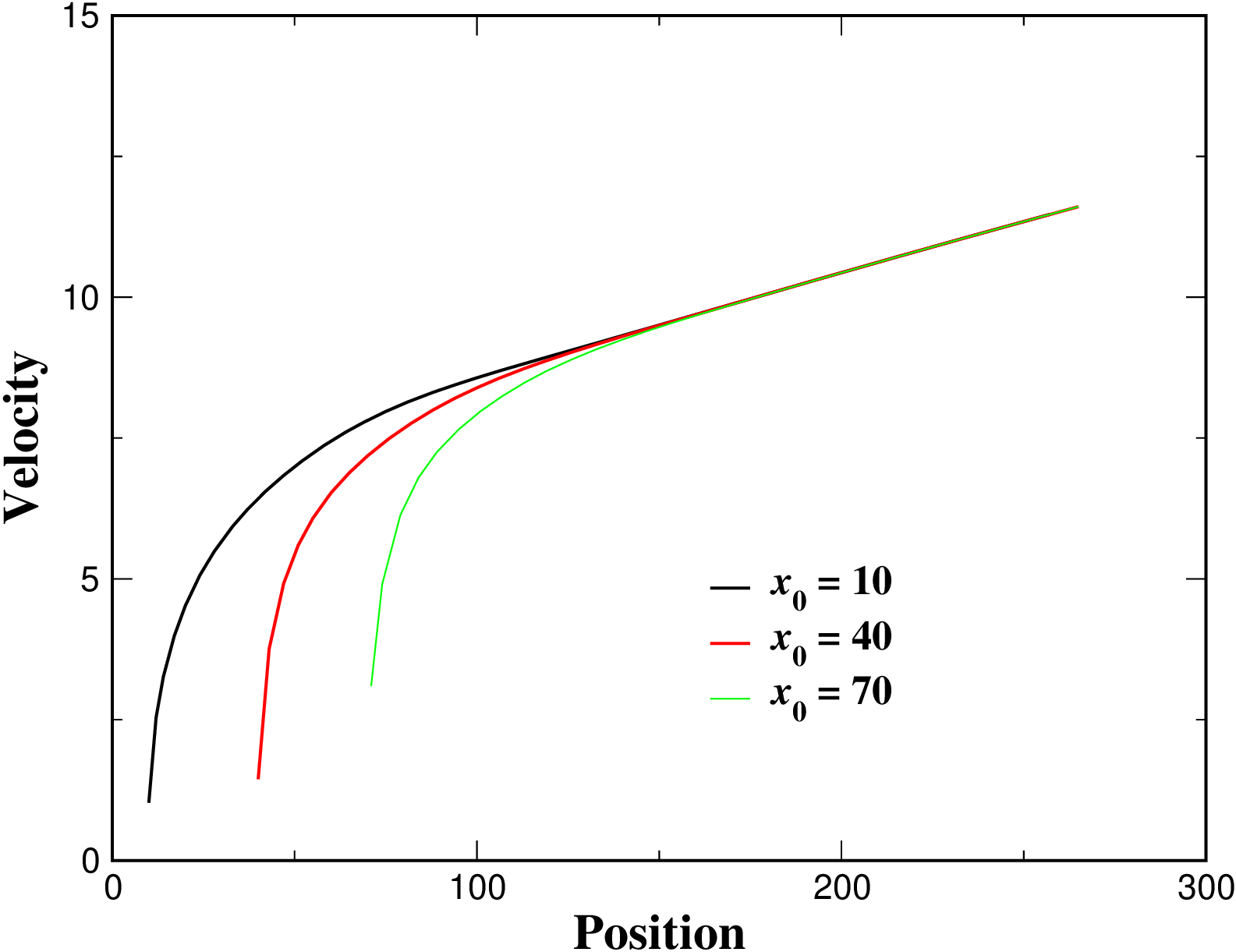}
  \includegraphics[width=.23\textwidth]{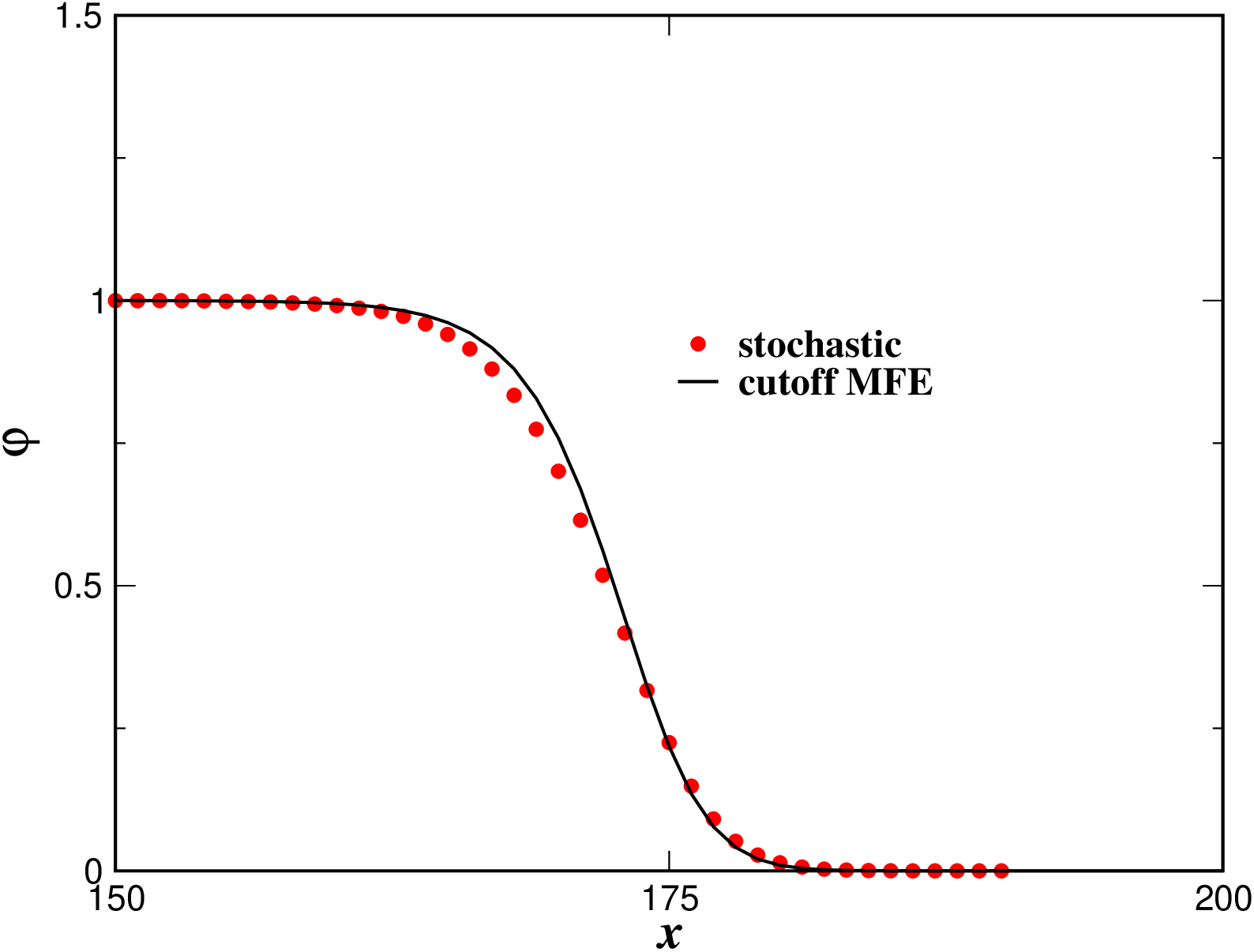}

\caption{ (a)  Regularity of the cutoff MFE; same parameters as Fig. 1a 
except for the cutoff of $10^{-6}$.  (b) Average profile from 
twenty simulations of the (quasistatic) stochastic system with 
$N=10^{6}$; also plotted is the corresponding cutoff MFE graph 
with $k=0.3$ }
\label{fig2}
\end{figure}

Given the highly unusual velocity results, an analytic treatment of
the cutoff MFE at large $N$ is clearly worthwhile. We restrict our
attention to the quasistatic case where the problem is a ``standard"
one of finding a homoclinic trajectory in the traveling coordinate
$z$. In what follows, it will become clear that spatial discretization
effects cannot be neglected and hence we keep the explicitly discrete
form of the hopping term, with lattice spacing $a$. The first key idea
is that the nonlinearity is important only near the bulk state
and in that region, diffusion is much less important than drift if the
velocity is large. In this range of sites, then, the full equation
\begin{eqnarray}
\label{full}
0 & = & D(\phi(z+a)+\phi(z-a)- 2\phi(z))  \nonumber \\ & & \; \; + v\phi'   +r(z)( \phi-\phi^2)\theta(\phi-k/N) \ .
\end{eqnarray}
can be solved by neglecting diffusion; this leads to
$$
\label{bulk}
\ln\left(\frac{1-\phi}{\phi}\right) = \left\{\begin{array}{cc}
    r_0(x + \alpha x^2/2)/v \quad\quad&  \\
    r_{\text{min}} (x-x_{\text{m}})/v + r_0(x_{\text{m}} + \alpha
    x_{\text{m}}^2/2)/v \quad\quad & \end{array}\right. 
$$
for $x \gtrless x_{\text{m}}$ respectively, where $x_{\text{m}}$ is the
point where the minimum reaction rate is reached, i.e.  $r_0+\alpha
x_{\text{m}}=r_{\text{m}}$.  We have chosen $\phi (0) =1/2$, 
to fix the translation invariance, which is a slightly different
definition of the front location, amounting to a small shift in $r_0$.

\begin{figure}
\includegraphics[width=.48\textwidth]{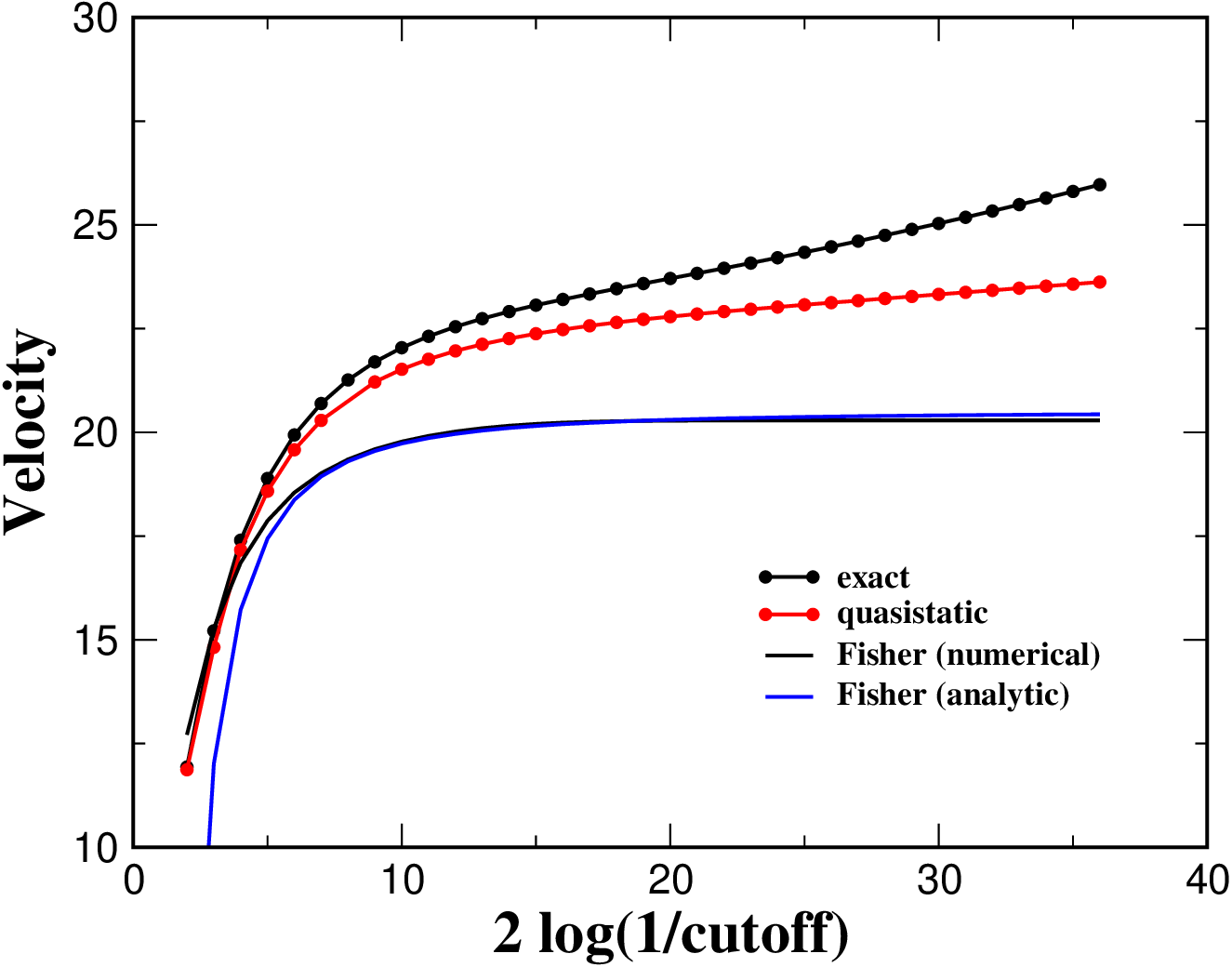}

\caption{Cutoff MFE equation for $D=5$, $r_0 =1$ and $\alpha =0.1$, velocity 
at $x=200$ versus $log N$. Shown for comparison is the absolute gradient 
model, the quasistatic model with $\tilde{r}_0 = r_0 +200 \alpha$, a 
numerical solution of a cutoff version of the usual Fisher equation with 
$r_f = r_0+200 \alpha$ and the Brunet-Derrida analytic approximation 
thereof.} 
\label{fig3}
\end{figure}

The above result needs to be matched to the solution near the cutoff
point, where the nonlinearity can be dropped. As initially suggested
by Rouzine, et al.~\cite{rouzine}, the resultant linear equation can be
approached via the WKB ansatz $\phi = e^{S(x)}$, 
with the key observation that the cutoff
point must occur close to the WKB turning point where $\phi$ begins to
oscillate.  This is the only way that there is enough freedom to do the
matching. The WKB equation takes the form
\begin{equation}
\label{wkb}
0 = \frac{4D}{a^2}\sinh^2(aS'/2) + vS' + r_0 + \alpha z \ .
\end{equation}
Already at this point, we get a nontrivial result.  We can
de-dimensionalize this equation by introducing $T = a/\ell S$, $y =
z/\ell$, $\ell = v/(a\alpha )$ so that the equation reads
\begin{equation}
\label{scaledwkb}
0 = \frac{4D}{va}\sinh^2(T'/2) + T' + y \ ,
\end{equation}
where the derivative is now w.r.t. $y$ and we have dropped the small term
$r_0 a/v$.\cite{r0-comment}  Thus, $S$ (i.e. $\ln N$, once
we do the matching) scales like $D/(\alpha a^3)$ times a function of
the dimensionless parameter $va/D$, so that the leading-order results for all $a$,
(for a given $\alpha$ and $D$) should lie on a universal curve.
Furthermore, we recover the idea already discussed in Ref.
(\cite{rouzine}) that $a$ is a singular perturbation as far as the large
velocity limit goes, since no matter how small $a$ is, the parameter
$va/D$ eventually goes to infinity.

Returning to Eq. (\ref{wkb}), the turning point is given by the 
discriminant equation
$\frac{dz}{dS'}=0$, yielding
\begin{equation}
0 = \frac{2D}{a} \sinh(aS'_*) + v \ ,
\end{equation}
which gives
\begin{equation}
S'_* = \ln\left(\sqrt{1+\frac{v^2 a^2}{4D^2}} - \frac{va}{2D}\right) \ .
\end{equation}
If we match to the bulk solution near $z=0$, $\phi$ declines by an
amount related to the change in $S$ from $z=0$ to the turning point
$z_*$.  This is given by
\begin{eqnarray*}
&S_* =& \int_0^{S'_*} dS'\,S' \frac{dz}{dS'}  \nonumber\\
&=& \int_0^{S'_*} \frac{dS'}{\alpha} \,S' [-(\frac{2D}{a}\sinh(aS') + v)] \nonumber\\
& = & -\frac{1}{\alpha} \left[\frac{2D}{a^2}S'_*\cosh(aS'_*) - \frac{2D}{a^3}\sinh(aS'_*) + \frac{v}{2}(S'_*)^2 \right] \ .
\label{lat-leading}
\end{eqnarray*}
In the geometrical optics approximation, $k/N \approx \phi(z_c) \approx \phi(z_*) = e^{S_*}$, giving us an our lowest-order answer for the velocity.
To improve upon this answer, one needs to both go beyond the geometrical
optics approximation, and to develop a connection formula to go past the
turning point.  The latter involves writing $\phi = e^{S'_* x} \psi$, 
where $\psi$ is smooth on the lattice scale~\cite{cohen-kess}, 
and showing that $\psi$
satisfies an Airy equation. This procedure will be described in detail in 
a future publication\cite{in-prep}, and here we just cite the results.
The most significant correction, we find, comes from the further 
exponential decay, at rate $S'_*$,
of $\phi$ between the turning point and the cutoff point, which is near the 
zero of the
Airy function solution to the $\psi$ equation in the connection region.  
This gives an additional contribution of
$-S'_* \xi_0 (k/(D\cosh(aS'_*)))^{-1/3}$ to $S$, where $\xi 0 =
-2.3381$ is the location of the first zero of the Airy function.  As
already mentioned, the solution for the velocity is just $\ln (N/k) = -S$,
so that the final expression becomes
\begin{eqnarray}
\ln (N/k) &=& \frac{1}{\alpha} \left[\frac{2D}{a^2}S'_*\cosh(aS'_*) - \frac{2D}{a^3} \sinh(aS'_*) +\frac{v}{2}(S'_*)^2 \right]  \nonumber \\ && - \xi_0 S'_* \left( \frac{\alpha}{D\cosh(aS'_*)}\right)^{-1/3}
\label{lat-correct}
\end{eqnarray}

Let us examine the various limits of the this expression.  First, in
the continuum limit, $av/D \ll 1$.  Then $S'_*=v/2D$, so that $aS' \ll
1$.  Then, $S_* = (2D(S'_*)^3/3 + v(S'_*)^2/2)/\alpha$; the correction
term is $-S'_* \xi_0 (k/D)^{-1/3}$. Combining these gives the formula
\begin{equation}
\label{asym_corrected}
v = 2(D^2 r_0 \alpha)^{1/3}\left[(3\ln (N/k))^{1/3}  + \xi_0(3\ln(N/k))^{-1/3}\right] \ .
\end{equation}
The scaling of $v$ as $\ln ^{1/3} N$ is an agreement with the results of Tsimring, et al. for a continuum evolution 
model~\cite{tsimring}.
Now, as we mentioned above, for finite $a$, $av/D$ is eventually large
for sufficiently large $N$. Then, $S'_* = -\ln(va/D)/a$.  This gives
$S_* = v/(a^2 k) (-\ln(va/D) + 1 + \ln^2(va/D)/2)$.  Now, for very
large $va/D$, $S_* \approx -v\ln(va/D)/(a^2 k)$.  However, this is
only valid for $ln(va/D) \gg 2$. In fact, it is a reasonable (20\%)
approximation only for $\ln(va/D)$ bigger than 10, so that $v$ would
be unreasonably large.  Thus, a strict asymptotic expansion is of no
use. Our analytic results based on equation \ref{lat-correct} are
compared to numerical solutions in Fig.4; more details are given
elsewhere~\cite{in-prep}. Note that the full expression fits the data
quite well, although the leading-order formula is not very accurate.
\begin{figure}
  \includegraphics[width=.23\textwidth]{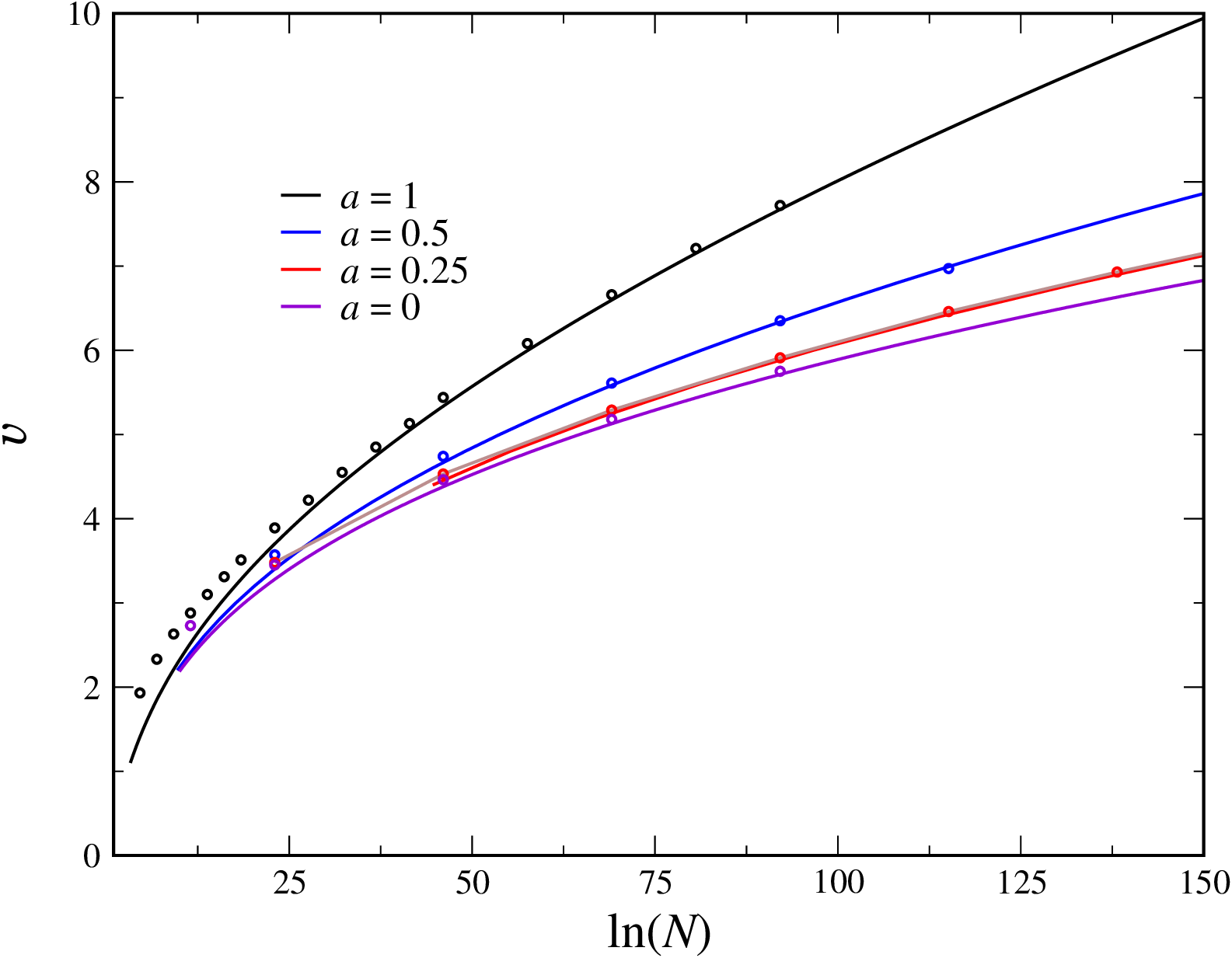}
  \includegraphics[width=.23\textwidth]{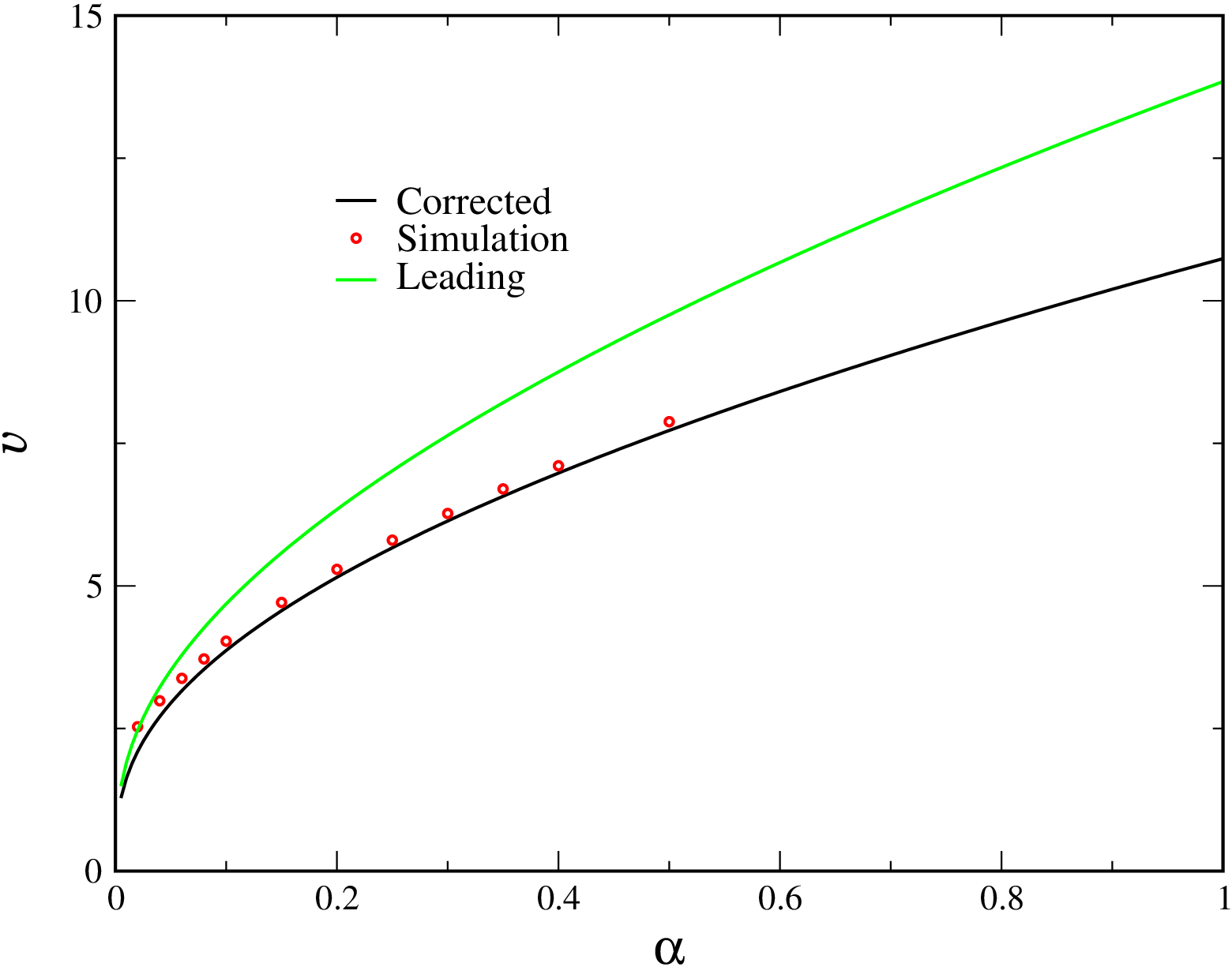}
\caption{(a) Velocity, $v$, vs. $\ln(N)$ for lattice spacings $a=1$, 0.5, 0.25, and 0,
  $\alpha=0.1$, $D=r_0=1$. from simulation, together with the analytic
  approximation Eq. (\protect{\ref{lat-correct}}) (b) Velocity, $v$, vs.
  $\alpha$ at $\ln N =25$, versus leading-order and full analytic
  expressions. In both figures, $k=1$.}
\label{fig_manya}
\end{figure}

One of the most interesting aspects of the above calculation is that
it does at all depend on form of the solution past the cutoff point;
the mere existence of a cutoff is enough to force the system to the
WKB turning point and hence fix the velocity. This is perhaps the
reason why the cutoff MFE mimics the actual stochastic model; the fact
that the region past the cutoff is in reality highly stochastic should
not alter the velocity fixation which occurs in the part of space
where fluctuation effects are indeed negligible. Turning this
qualitative argument into a full theory remains a challenge for future
work.

In summary, we have introduced a new class of fronts in
reaction-diffusion systems and showed how fluctuations must be taken
into account, even if only heuristically, if one wished understand
their behavior. Using a cutoff MFE approach, we can understand in
detail the anomalous velocity behavior, at least for the quasistatic
case. Further work should address the role of acceleration and front
stability.

The work of HL has been supported in part by the NSF-sponsored Center
for Theoretical Biological Physics (grant numbers PHY-0216576
and PHY-0225630).


\begin{thebibliography}{99}
\bibitem{stochastic} The simulation is performed by alternating
  binomial processes for hopping (with probability per particle $Ddt$ in
each direction)
  with the reaction probability per B particle equal to
  $1-\left(1-\frac{r\,dt}{N}\right)^{N_A}$. This form guarantees that the
  probability never exceeds unity and leads to the mean-field equation
  $\rho _t \simeq \rho (1- \rho /N) +D \rho ''$. Setting $\phi =\rho
  /N$ leads to the Fisher equation.
\bibitem{r0-comment} The scaling derived here assumes $\tilde{r}_0$ of
  O(1) and hence gives a velocity constant in time even when we choose
  $\tilde{r}_0 = r_0 +\alpha t$ in the quasistatic approximation to
  the absolute gradient case. This will break down at large times. One
  can derive thereafter a new scaling assuming large $\tilde{r}_0$ of
  $O(v)$; this will be presented elsewhere~\cite{in-prep}.
\end{thebibliography}
\end{document}